%Paper: hep-th/9406199
%From: khalpern@MIT.EDU
%Date: Wed, 29 Jun 94 14:15:09 EDT

%%%%%%%%%%%%%%%%%%%%%%%%%%%%
%
%   This paper is in regular tex.  It has a uuencoded 
%   file appended at the end.  This figure must be reproduced
%   on its own.  Instructions for doing so are contained in its
%   header.
%%%%%%%%%%%%%%%%%%%%%%%%%%%%

\magnification=1200
\baselineskip=24pt plus 2pt minus 2pt
\tolerance=10000
\hfuzz=10pt

\countdef\eqnum=11    % Number is dummy
\eqnum=0              % This will number equation from 1.
% Equation numbering macros:
%  To number an equation automatically, use \e instead of \eqno
%  To name an equation {junk}, place after the equation \enam{\junk}
%  Refer to equation as \junk. The equation number will be printed.
\def\e{\global\advance\eqnum by 1 \eqno(\the\eqnum)}
\def\enam#1{\xdef#1{(\the\eqnum)\ }}

\nopagenumbers       
\line{\hfil                                    CTP \#2281}
\line{\hfil                                    June, 1994}
\vfill
\vskip 0.5 truein
\centerline{Fixed-Point Structure of Scalar Fields}
\vskip 0.3 truein
\centerline{Kenneth Halpern and Kerson Huang}
\vskip 0.3 truein
\centerline{Department of Physics}
\centerline{and Center for Theoretical Physics, Laboratory for Nuclear Science}
\centerline{Massachusetts Institute of Technology, 
Cambridge, Massachusetts, MA 02139}
\vskip 0.5truein
\centerline{                ABSTRACT}
\par\noindent

We search for alternatives to the trivial $\phi^4$ field theory
by including arbitrary powers of the self-coupling.  Such theories are 
renormalizable when the natural cutoff dependencies of the coupling
constants are taken into account.  
We find a continuum of fixed points, which includes the well-known
Gaussian fixed point. The fixed point density has a maximum at a location 
corresponding to a theory with a Higgs mass of approximately 2700 GeV.  
The Gaussian fixed point is UV stable in some directions in the 
extended parameter space.  Along such directions we obtain non-trivial
asymptotically free theories.  

\vfill\noindent
PACS 11.10.Gh, 12.15.Cc, 64.60.Ah
\eject
\pageno=1                        %set next page to be page 1
\footline={\hss\tenrm\folio\hss} %restore page numbering

\noindent
{\bf 1. Introduction and Summary}

The Higgs field of the standard model, usually taken to be a 
scalar field with quartic self-interaction,
has been beclouded by the issue of ``triviality''; namely,
the renormalized coupling vanishes in the limit of infinite cutoff [1]. 
This has been verified by numerical calculations [2], and the
implications for phenomenology have been examined [3].
In terms of Wilson's renormalization group (RG) [1], 
the reason for triviality is that the continuum (or infinite-cutoff) 
limit is identified with an infrared (IR) Gaussian fixed point, so that
the theory approaches a free field theory in the low-energy limit.
It is natural to ask whether there are alternative continuum limits
that yield a non-trivial theory. As one of us [4] noted earlier, in a
 $\phi^4$ theory in $4-\epsilon$ dimensions there are RG trajectories on which
the Gaussian fixed point appears as an ultraviolet (UV) fixed point. 
Theories built along these trajectories would be non-trivial 
and  asymptotically free. Unfortunately, they become trivial as 
$\epsilon\to 0$. However, this scenario has led us to search for
non-trivial theories in an extended parameter space. 
In this note we report on some positive results.

The model being investigated is an $N$-component real scalar field theory
in $d$ dimensions, with arbitrary power-law self-coupling. We are interested
in how the couplings transform under a change of energy scale, and
the inclusion of all powers is necessary for closure under RG.
The theory remains renormalizable
in the usual perturbative sense when we recognize that the 
coupling constants must depend on the cutoff in specific manners.
These dependencies are such that the S-matrix of the theory for $d=4$ 
is that of an effective $\phi^4$ theory, whose effective
coupling depends on all the coupling constants of the underlying theory.
To study the renormalization of the effective coupling, 
we must examine the RG flow in the infinite-dimensional parameter space
of the extended theory. This is done using Wilson's method 
of momentum-shell integration, and the main results
may be summarized as follows:

(a) In the extended parameter space, some RG trajectories flow into the
Gaussian fixed point, while others flow out of it. The Gaussian fixed 
point is UV-stable with respect to the latter type of
trajectories, along which the theory is 
non-trivial and asymptotically free. Spontaneous symmetry breaking occurs
along some trajectories of this type.

(b) There exists a one-parameter continuum of non-trivial fixed points. For 
the Higgs field ($N=4, d=4$) 
the density of fixed points is maximum at a location corresponding 
to a potential with broken symmetry. The Higgs mass calculated from this 
potential is approximately 2700 GeV. 
\bigskip
\noindent
{\bf 2. The Model}

The Euclidean action of our model is
$$\eqalign{
S[\phi]
&= \int d^dx \left[ {1\over 2}(\partial\phi_{i})^2 + V(\phi^2) \right]\cr
V(\phi^2)&=\sum_{n=1}^\infty g_{2n} (\phi^2)^n\cr
}                                \e
$$
\enam{\action}
where $\phi_i(x)\ (i=1,\cdots, N)$ are real fields, and $\phi^2=\phi_i\phi_i$.
There is a momentum cutoff (or equivalent inverse lattice spacing) 
$\Lambda$, which is assumed to be the only intrinsic scale in the theory.
Since the coupling constant $g_K$ has dimension (momentum)$^{K+d-Kd/2}$,
it must depend on the cutoff according to
$$
g_K=u_K\Lambda^{K+d-Kd/2}	\e
$$
\enam{\coupling}
where $u_K$ is a dimensionless parameter. We define $u_2=r/2$, and write for
the bare mass square 
$$
m_0^2\equiv 2 g_2 =r\Lambda^2    \e
$$
\enam{\baremass}

In the conventional renormalization scheme, a dimensionless 
coupling constant is considered renormalizable; 
but any higher coupling is rejected
on the basis that it leads to divergencies that cannot be cancelled by
a finite number of counter terms. In $d$ dimensions,
according to this rule, only $\phi^M$ and lower-order theories 
are renormalizable, where $M=2d/(d-2)$.
With the cutoff dependence {\coupling} taken into account, 
however, the higher couplings do not generate new divergencies, 
and are renormalizable. The reason is that they
vanish in the limit $\Lambda\to\infty$, thus supplying extra
convergent factors to Feynman graphs [5]. A more detailed analysis,
supplementing the usual power counting [6] with the cutoff dependencies
from {\coupling}, shows that, for $d\ge 4$,
a higher vertex can contribute to
a skeletal graph only if the graph has $M$ or fewer external lines.
For a non-skeletal graph with more than $M$ external legs, higher
vertices can contribute only via vertex insertions. This
means that the S-matrix of the theory is that of 
an effective $\phi^M$ theory, whose effective coupling depends on
all the dimensionless couplings $u_K$. Thus,
renormalization must be discussed  in the context
of the underlying theory. In particular,
the fixed points of the effective coupling are projections of 
fixed points in the extended parameter space,
and, except for the Gaussian fixed point, cannot be found from the 
effective theory. 
\bigskip
\goodbreak
\noindent
{\bf 3. Renormalization-Group Equations }
\nobreak

We make use of Wilson's RG transformation [1], 
which thins out degrees of freedom by 
(a) integrating out Fourier components of $\phi_i$ with momenta 
lying in a shell between $\Lambda$ and $\Lambda/b$, 
(b) rescaling the cutoff back to $\Lambda$, and (c) rescaling
$\phi(x)$ to restore normalization in \action. We put
$$
t=\ln b								\e
$$
\enam{\scale}
so that $t=0$ corresponds to the energy scale at the cutoff.
The transformation is effected by first decomposing the field into two terms 
$\phi=\phi_s + \phi_f$,
where $\phi_s$ (the slow piece) has non-vanishing Fourier components
only for momenta $k$ with $|k|<\Lambda/b$, while $\phi_f$ (the fast piece)
has $\Lambda/b\le |k| \le\Lambda$. The action is split into two terms,
the free-field (quadratic) action $S_0[\phi]$, which includes the kinetic 
and the bare-mass term, and $S_1[\phi]$, which contains the remaining terms.
We can write $S[\phi_s+\phi_f]=S_0[\phi_s]+S_0[\phi_f]+S_1[\phi_s+\phi_f]$, 
and put the partition function in the form
$$\eqalign{
Z&=\int (D\phi_s) \exp\{-S_0[\phi_s]\}\int (D\phi_f)\exp\{-S_0[\phi_f] 
- S_1[\phi_s+\phi_f] \}\cr
&\equiv 
\int (D\phi_s) \exp\{-S'[\phi_s] \}\cr
}\e
$$
\enam{\partition}
We then rescale all momenta by a factor $b$ to restore the apparent cutoff
to $\Lambda$, rescale $\phi_{s}$ to restore our former normalization, and read off the new coupling constants from the rescaled $S'$.
The RG trajectories in the parameter space are generated through
successive RG transformations.

In practice the RG transformation is carried out in perturbation theory by
expanding the partition function in powers 
of $S_1[\phi_s+\phi_f]$, where $\phi_f$ is the field to be integrated over,
with $\phi_s$ held fixed. In the Feynman graphs of
this expansion, all internal lines correspond to $\phi_f$, and
external lines correspond to $\phi_s$. The renormalized value of $u_K$ has 
contributions from connected graphs with $K$ external lines.

We calculate only to first order in $t$, with an infinitesimally thin
momentum shell. This yields $\partial u_K / \partial t$ at $t=0$, or
$\beta$-functions at $t=0$. For this calculation we only need
to include tree and one-loop graphs, because the momenta of internal lines
have only an infinitesimal range, and any additional loop 
integration will vanish in the limit $t\to 0$, because its range goes to zero.
In general, the RG transformation
generates all powers of the self-coupling, even if they were not
present in the beginning. In addition, derivative couplings will
arise, but are ignored in the current analysis.

The continuum limit $\Lambda\to\infty$ is not trivial, 
because $\Lambda$ is invisible: In the action, $\Lambda$ can be absorbed
through a rescaling of the field. To find the value of $\Lambda$,
we have to compute some physical quantity, for example 
a correlation length, which will be given in units of $\Lambda^{-1}$
(or equivalent lattice spacing.) We only know that $\Lambda\to\infty$ 
when the correlation length diverges, and this can be true only at 
certain RG fixed points. A continuum limit therefore
corresponds to some fixed point, and different 
fixed points define different physical theories. 

As mentioned above, the $\beta$-functions are calculated at $t=0$ 
(the energy scale of the cutoff)
by summing one-loop graphs.  But, since 
$\Lambda$ does not explicitly appear in the action, these functions are
characterized solely by the values of the $u_K$. Thus, the
$\beta$-functions are functions only of the $u_K$, and 
our one-loop RG equations are exact except for the neglect of derivative
coupling terms. They are as follows:
$$\eqalign{
{\partial u_{2n}\over \partial t} &= (2n+d-nd) u_{2n}
+{S_d\over 2}\sum_{k=1}^n { (-1)^{k-1} 2^k\over k(1+r)^k }
\mathop{{\sum}'}_{\{m_1,\cdots,m_k\}} 
P(m_1,\cdots,m_k) u_{2m_1}\cdots u_{2m_k}\cr
(n&=1,\cdots, \infty)\cr
}\e
$$
\enam{\RG}
where $u_K=u_K(t)$, and
$$
P(m_1,\cdots,m_k)=(m_1\cdots m_k) \{ [(2m_1-1) \cdots (2m_k-1)] +(N-1)\} \e
$$
In \RG, $r$ is the bare-mass parameter defined in \baremass, and
$S_d=2^{1-d}\pi^{-d/2}/\Gamma(d/2)$ is the surface
area of a unit $d$-sphere
divided by $(2\pi)^d$, with $S_4=(8\pi^2)^{-1}$. 
In the sum over $\{m_1,\cdots,m_k\}$, the prime indicates the 
conditions
$$\eqalign{
m_i&=2,\cdots,n-k+2\cr
\sum_{i=1}^k m_i&=n+k\cr} \e
$$
On the right side of \RG, the first term comes from
rescaling. The second term contains the one-loop contributions, in which
$k$ is the number of vertices on the loop, with respective orders 
$2m_1,\cdots, 2m_k$.
\bigskip
\noindent
{\bf 4. Gaussian Fixed Point}

An obvious fixed point of the RG equations is the Gaussian fixed point, 
with all $u_{2n}=0$. The directions of stability and instability can
be found by diagonalizing the linearized RG equations.  
A negative eigenvalue corresponds to an IR-stable direction, 
while a positive one corresponds to an UV-stable direction. 

We quote the results for the case
$d=4, N=4$, which  corresponds to the Higgs field. For the eigenvalue
$$
\lambda=2(a+1) \e
$$
the components of the eigenvector are given by
$$
u_{2n}(0) = {r(0) (8\pi^2)^{n-1}
[ (a+1)\cdots(a+n-1) ]\over n! (n+1)! } \quad(n=1,\cdots,\infty)\e
$$
The Gaussian fixed point is IR-stable
for $a<-1$, UV-stable for $a>-1$, and marginal for $a=-1$. 
The eigenpotential defined by 
$V_a(\phi^2,t)\equiv\sum_{n=1}^\infty u_{2n}(t)(\phi^2)^n$ has the property
$$
(\partial V_a/\partial t)_{t=0}=\lambda V_a	\e
$$
and is given at t=0 by
$$
V_a(\phi^2,0)  =
{r(0)\over 8\pi^2 a} [ M(a,2,8\pi^2\phi^2)-1]	\e
$$
where $M(a,b,z)$ is a Kummer function [7]:
$$
M(a,b,z)=\sum_{n=0}^\infty{ a (a+1)\cdots (a+n-1) \over b (b+1)\cdots (b+n-1)}
{z^n \over n!}
={\Gamma(b)\over \Gamma(b-a)\Gamma(a)}\int_0^1 dt \ e^{zt} t^{a-1} 
(1-t)^{b-a-1}	\e
$$
The series breaks off to become a polynomial for negative integer values
$a=-1,-2,\cdots$. The case $a=-1$, corresponding to the free field,
is marginal (has eigenvalue zero.) The cases $a=-2,-3,\cdots$ are IR-stable.
This shows that all polynomial interactions become irrelevant in the 
low-energy limit, and the higher the polynomial 
degree the greater the irrelevancy. This justifies the neglect of
higher terms than $\phi^4$ in the conventional Higgs sector.
There also exist other IR-stable cases corresponding to non-polynomial
interactions with non-integer $a<-1$.

All the positive eigenvalues, with $a>-1$, correspond to non-polynomial 
interactions. The eigenpotentials in these cases rise
like $z^{a-2} e^z$ for large $z\equiv 8\pi^2\phi^2$, and thus have
different norms from the polynomial potentials. They describe non-trivial 
asymptotically free theories. 
The ``running'' potential $V_a(\phi^2,t)$ can be found by solving the
non-linear RG equations with given initial conditions.
Initial potentials with $0<\lambda<2$ and $r(0)<0$ yield theories with
asymptotic freedom and spontaneous symmetry breaking.
Sufficiently closed to the Gaussian fixed point, where the linear
approximation holds, the eigenpotenials corresponding to these conditions
have one and only one minimum. An approximate calculation of properties 
near the minimum gives the following estimate for the squared
ratio of Higgs mass to vacuum field:
$$
{m^2\over \langle\phi^2\rangle} \approx {8\pi^2\over 3}\sqrt{2\lambda(1
+\lambda)}(-r(t)) \e
$$
\bigskip
\goodbreak
\noindent
{\bf 5. Non-Trivial Fixed Points}
\nobreak

To find all the fixed points of our theory, we set
$\partial u_{2n}/ \partial t =0$ in {\RG}, thereby obtaining 
a recursion formula for the $u_{2n}$ at the fixed point, with  
one free parameter $r$.  Therefore, we have a one-dimensional
continuum of fixed points --- a fixed line --- with the
Gaussian fixed point located at $r=0$.  Fig.1 shows a plot of $u_{2n}$
as a function of $r$ for $n=2\cdots 30$, for the case $d=4, N=4$.
 They are normalized to 1 at the maxima. With the
exception of the case $n=2$, all maxima occur at approximately the
same point: $r_{\rm max}=-0.65$. This indicates that the density of fixed 
points has a maximum at $r=r_{\rm max}$. 
The potential at this point exhibits spontaneous 
symmetry breaking, and yields a ratio of Higgs mass to vacuum field
of approximately 10. Using a vacuum field of 273 GeV gives a 
Higgs mass of approximately 2700 GeV.
The significance of the accumulation point on the fixed line is
yet unclear, and merits study.  

Further results and details of calculations will be reported in a 
separate publication. This work is supported
in part by funds provided by the U.S. Department of Energy under
contract \# DE-AC02-76ER03069, and cooperative agreement \#
DE-FC02-94ER40818.
\vfill
\eject
\centerline{References}
\item{[1]} K.G. Wilson and J.Kogut, Phys. Rep. {\bf C12}, 75 (1974).
\medskip
\item{[2]} K. Huang, E. Manousakis, and J. Polonyi, Phys. Rev., {\bf 35}
(1987) 3187; J. Kuti, L. Lin, and Y. Shen, Phys. Rev. Lett., {\bf 61}
(1988) 678; and earlier references therein.
\medskip
\item{[3]} R. Dashen and H. Neuberger, Phys. Rev. Lett., {\bf 50} (1983) 1897.
\medskip
\item{[4]} K. Huang, ``An Asymptotically Free $\phi^4$ Theory,''
MIT CTP preprint 2208 (1994), unpublished.
\medskip
\item{[5]} J. Polchinski, Nucl. Phys. {\bf B 231}, 269 (1984).
\medskip
\item{[6]} See, for example, K. Huang, {\it Quarks, Leptons, and Gauge Fields},
2nd Ed. (World Scientific, Singapore, 1992), p.188.
\medskip
\item{[7]} M. Abramovitz and I.A. Stegun, {\it Handbook of  
Mathematical Functions} (National Bureau of Standards, Washington, 1964),
p.503.

\bigskip
\centerline{\bf Figure Caption}
\bigskip\noindent
Fig.1.  Plot of the coefficients $u_{2n} (n=2\cdots 30)$ 
along the fixed line parametrized by $r$, normalized to 1 
at the maximum of each curve.
With one notable exception  ($n=2$) 
all the maxima approximately coincide at $r=-0.65$.
\end
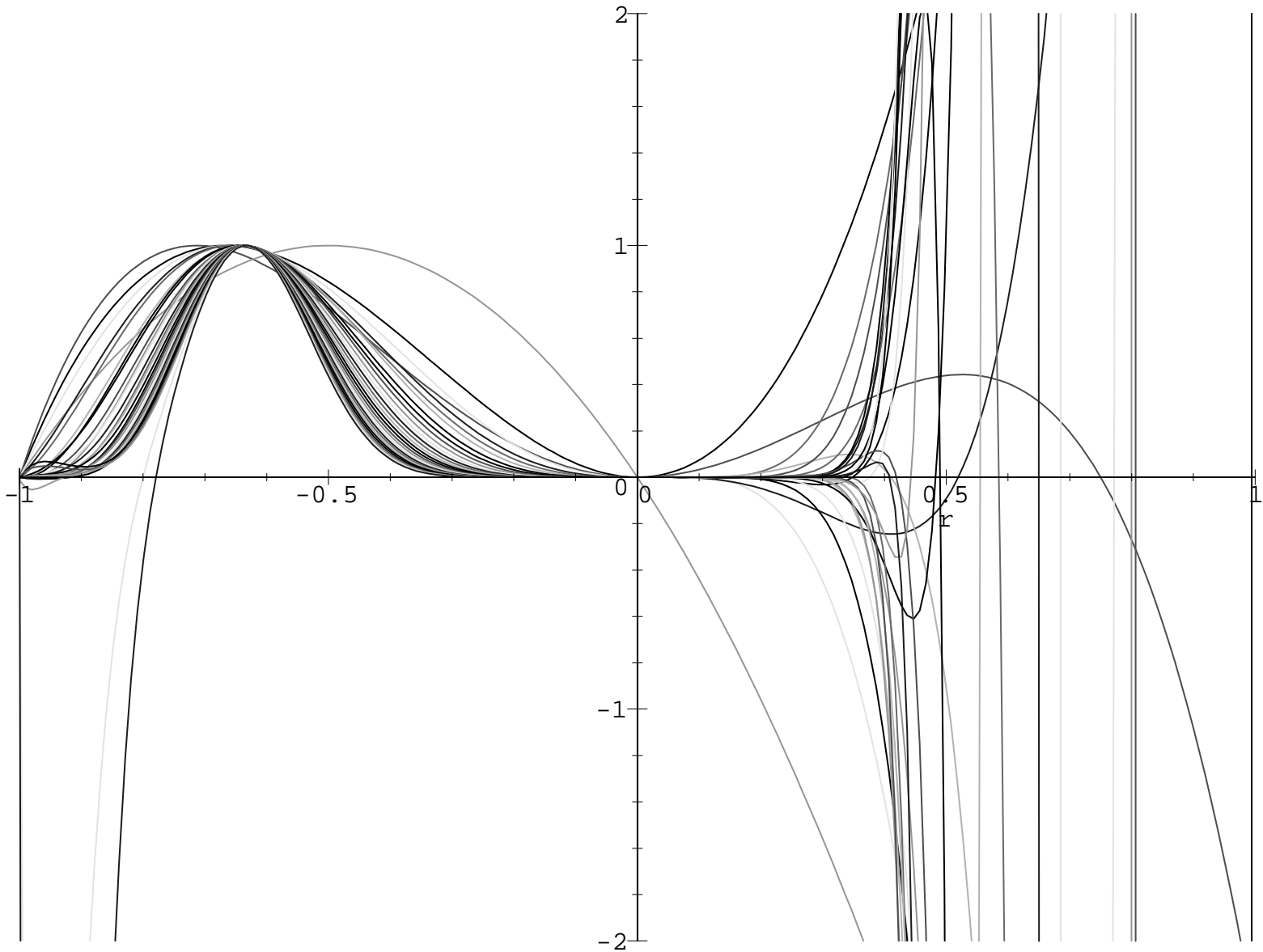
\end